# Exotic magnetic behaviour and evidence of cluster glass and Griffiths like phase in Heusler alloys $Fe_{2-x}Mn_xCrAl$ ($0 \leq x \leq 1$)

Kavita Yadav[1], Mohit K. Sharma[1], Sanjay Singh[2], and K. Mukherjee[1]

[1]School of Basic Sciences, Indian Institute of Technology Mandi, Mandi 175005, Himachal Pradesh, India

[2]School of Materials Science and Technology, Indian Institute of Technology (Banaras Hindu University), Varanasi, 221005, India

## Abstract

We present a detailed study of structural, magnetic and thermodynamic properties of a series of Heusler alloys $Fe_{2-x}Mn_xCrAl$ (x=0, 0.25, 0.5, 0.75 and 1). Structural investigation of this series is carried out using high resolution synchrotron X-ray diffraction. Results suggest that with increasing Mn concentration, the $L2_1$ structure of $Fe_2CrAl$ is destabilized. The DC magnetization results show a decrement in paramagnetic (PM) to ferromagnetic (FM) phase transition temperature ($T_C$) with increasing Mn concentration. From the systematic analysis of magnetic memory effect, heat capacity, time dependent magnetization, and DC field dependent AC susceptibility studies it is observed that, $Fe_2CrAl$ exhibits cluster glass(CG)-like transition approximately at 3.9 K ($T_{f2}$). The alloys, $Fe_{1.75}Mn_{0.25}CrAl$ and $Fe_{1.5}Mn_{0.5}CrAl$ exhibit double CG-like transitions near $T_{f1}$~22 K, $T_{f2}$~4.2 K and $T_{f1}$~30.4 K, $T_{f2}$~9.5 K respectively, however, in $Fe_{1.25}Mn_{0.75}CrAl$, a single CG-like transition is noted at $T_{f2}$~11.5 K below $T_C$. Interestingly, FeMnCrAl shows the absence of long ranged magnetic ordering and this alloy undergoes three CG-like transitions at ~ 22 K ($T_f^*$), 16.6 K ($T_{f1}$) and 11 K ($T_{f2}$). At high temperatures, a detailed analysis of temperature response of inverse DC susceptibility clearly reveals the observation of Griffiths phase (GP) above 300 K ($T^*$) in $Fe_2CrAl$ and this phase persists with Mn concentration with a decrement in $T^*$.

## Introduction

In the past few decades, Heusler alloys have received considerable attention due to a variety of novel physical properties like magneto-optical effect [1], magnetocaloric effect [2], unconventional superconductivity [3], half metallicity [4], magnetic shape memory effect [5], barocaloric effects [6], magnetoresistance [7], anomalous Hall Effect [8] and large exchange bias [9] exhibited by them. Ternary Heusler alloys are represented by the formula $X_2YZ$ where X and Y are transition metals and Z is usually *sp* element from III-VI groups of the periodic table. These systems generally crystallize in $L2_1$ structure with *Fm-3m* space group [1-5]. Among several compositions, $Fe_2Y'Al$ (Y′: 3d transition metals such as V, Cr and Ti) have been extensively investigated in recent years because considerable changes have been observed near the Fermi level in their electronic band structure [10]. For example, $Fe_2VAl$ show semiconducting behaviour in spite of metallic constituents and is nonmagnetic down to 2 K, inspite of the presence of Fe [11]. Also, $Fe_2TiAl$ exhibits optical and electronic properties similar to metals but has a low magnetic moment [10, 12]. On the other hand, it has been reported that this observed non-metallic character can be due to magnetic disorder. Interestingly, the ratio of Y′/Fe can be tuned to study the role of anti-site/chemical disorder on magnetic as well as transport properties. R. Saha *et al.* [13] investigated the magnetic properties of $Fe_2V_{1-x}Cr_xAl$ (x=0 to 1) and found that increment in Cr content promotes site disorder and alters the magnetic properties of $Fe_2VAl$.

Irrespective of elements present in Heusler alloys, there are significant chances of formation of the disordered structure as they show greater sensitivity towards the environment as well as elemental substitution. One such studied alloy is $Fe_2CrAl$. From first principles calculations it has been predicted to be a half metallic ferromagnet (HMF) system while, experimental investigations deviate from the theoretical results [14]. It has been reported that this alloy crystallizes in *B*2 type structure and exhibits semiconducting behaviour. It undergoes ferromagnetic transition with $T_C$ ~ 234 K [15]. Also, through band structure calculations, it is predicted that $Mn_2CrAl$ is a HMF with 100% spin polarization [16]. However, there are no experimental reports available in the literature till now. Hence, it will be interesting to explore the effect of Mn substitution at Fe-site in tuning the structural as well as magnetic properties of the $Fe_2CrAl$ alloy.

In this manuscript, we investigate the structural, magnetic and thermodynamic properties of a series of Heusler alloys $Fe_{2-x}Mn_xCrAl$ (x = 0, 0.25, 0.5, 0.75 and 1). Investigations are carried out through high resolution synchrotron X-ray diffraction, temperature and field dependent magnetization, time dependent magnetization, magnetic memory effect, DC field dependent AC susceptibility along with heat capacity studies. Our studies reveal that: i) increment of Mn at Fe site in $Fe_2CrAl$ destabilizes the $L2_1$ structure ii) the parent compound $Fe_2CrAl$ undergoes paramagnetic (PM) to ferromagnetic (FM) phase transition near $T_C$ ~202 K, followed by a cluster glass (CG)-like state near $T_{f2}$~3.9 K iii) with increment in Mn concentration, $T_C$ is suppressed to 120, 48 and 27 K for $Fe_{1.75}Mn_{0.25}CrAl$, $Fe_{1.5}Mn_{0.5}CrAl$ and $Fe_{1.25}Mn_{0.75}CrAl$ respectively while it is observed that $Fe_{1.75}Mn_{0.25}CrAl$ and $Fe_{1.5}Mn_{0.5}CrAl$ exhibit double CG-like transitions near $T_{f1}$ ~22 K, $T_{f2}$~4.2 K and $T_{f1}$~30.4 K, $T_{f2}$~9.5 K respectively iv) only one CG-like transition is observed near $T_{f2}$~11.5 K for $Fe_{1.25}Mn_{0.75}CrAl$ v) long ranged magnetic ordering is suppressed in FeMnCrAl and this alloy undergoes CG-like transitions at $T_f^*$~ 22 K, $T_{f1}$~16.6 K and $T_{f2}$~11 K vi) in the parent compound Griffiths phase-like singularities is noted and this is shifted down in temperature with an increase in Mn concentration. In this series of alloys, significant suppression in value of $T_C$ can be due to the presence of AFM coupling between Fe and Mn. Also, observation of more than one CG phase can be due to enhancement in magnetic anisotropy in the system, which dissociates the infinite clusters into smaller ones. In the present case, these different sized clusters freeze at a different temperature depending upon their anisotropic fields i.e. smaller and highly anisotropic clusters freeze near $T_{f2}$ whereas bigger and less anisotropic clusters freeze near $T_{f1}$ and $T_f^*$. Additionally, GP-like behaviour is noted in high temperature regime and such feature is observed due to the formation of short range ordered clusters which arise due to quenched/anti-site disorder between Fe and Al.

**Results**

**Structural analysis**

The Le-Bail refinements of room temperature XRD patterns of the $Fe_{2-x}Mn_xCrAl$ ($0 \leq x \leq 1$) alloys are shown in the Fig.1. All the reflections present in the XRD patterns could be identified using a cubic structure (*Fm-3m* space group). This confirms that all the samples are in single phase without any impurity. The refined lattice parameters and cell volume for all compositions are listed in table 1. It can be noted from the table 1 that the lattice parameter and hence the unit cell volume increases with increasing concentration of Mn. A comparison of the x-ray diffraction patterns of the parent and doped alloys (inset (b) of Fig. 1) reveals a gradual shift of diffraction lines to the lower angle side, thereby establishing that the dopant goes to the respective site. It has been suggested in literature that from the intensity ratio of (200) and (220) reflections i.e. $I_{200}/I_{220}$ one can determine the degree of site ordering of Fe and Al atoms [13]. Here, the intensity of (200) is low (as compared to the most intense peak) to be seen clearly. Hence we have plotted it for different alloys (in inset of fig. 1 (b)) along with variation in its Mn concentration. In $Fe_{2-x}Mn_xCrAl$, replacing Fe with Mn reduces the intensity of (200) reflection and hence the $I_{200}/I_{220}$ ratio decreases with Mn, which indicates enhancement in site disorder as shown in inset of fig.1 (c). This disorder destabilizes the $L2_1$ structure and can also alter the physical properties of the Heusler alloys [13, 17-20]. Therefore, in order to understand the effect of chemical/anti-site disorder on magnetic and thermodynamic properties of this series of alloys, magnetic and heat capacity studies have been carried out.

**Temperature and magnetic field dependent DC magnetization studies**

Temperature response of zero-field cooled (ZFC) and field-cooled (FC) magnetization of the series $Fe_{2-x}Mn_xCrAl$ ($0 \leq x \leq 1$) at 100 Oe are represented in Fig. 2 (a-e). It is observed that the parent compound $Fe_2CrAl$ undergoes magnetic phase transition near 202 K, which is confirmed through d($M(T)$)/d$T$ vs. $T$ plot (as shown in Fig. 2(f)). This magnetic transition can be attributed as PM to FM phase transition which is in accordance with the literature [15]. Interestingly, a weak bifurcation between ZFC and FC curves is also noted in low temperature regime. It is found that increment of Mn concentration in the parent compound suppresses this magnetic transition temperature to 120, 48 and 27 K for $Fe_{1.75}Mn_{0.25}CrAl$, $Fe_{1.5}Mn_{0.5}CrAl$ and $Fe_{1.25}Mn_{0.75}CrAl$ respectively. Also, in contrast to the parent compound, a large bifurcation between ZFC and FC magnetization curves is observed in low temperature region of $Fe_{1.75}Mn_{0.25}CrAl$ and $Fe_{1.5}Mn_{0.5}CrAl$. In all these alloys, one can see that the ZFC magnetization curves increase down to a particular temperature but show a downturn at low temperatures, signifying a possible presence of another magnetic phase. This downturn along with bifurcation in ZFC and FC curves is generally attributed to the presence of glassy magnetic phase or due to short range magnetic correlation or owing to the presence of superparamagnetic (SPM) phases and can also originate due to magneto-crystalline anisotropy and/or anti-site disorder [21]. Additionally, in high temperature regime of $Fe_{1.5}Mn_{0.5}CrAl$, i.e. in the PM phase a weak feature (deviation of ZFC curve) is noted which is also present in $Fe_{1.25}Mn_{0.75}CrAl$ and FeMnCrAl as illustrated in insets of Fig. 2(c), (d) and (e) respectively. This high temperature magnetic phase will be discussed further in section "Evidence of the presence of the Griffiths phase in the high temperature region". The isothermal magnetization curves as a function of the magnetic field at 2 K and 300 K for all the alloys are shown in Fig. 3 (a) and (b)). At 2 K, it is observed that the nature of the curves remains unchanged across the series and all the curves show insignificant hysteresis. It is also observed that there is a significant reduction in magnetization value obtained at 50 kOe for FeMnCrAl as compared with $Fe_2CrAl$. However, at 300 K, it is found that $Fe_2CrAl$ and $Fe_{1.75}Mn_{0.25}CrAl$ exhibit non-linear behaviour rather than expected linear behaviour of a pure PM phase, indicating the presence of some short range magnetic correlations above $T_C$.

**Role of magnetic anisotropy in low temperature regime**

In order to investigate whether magneto-crystalline anisotropy affects the nature of ZFC and FC curves below ordering temperature, we have analysed the magnetic field dependent magnetization response, using random anisotropy theory (RAT) developed by Chudnosky *et al.* [22]. In this theory, the magnetic ground state of an alloy can be determined by the strength of random anisotropy field, leading to the prediction of different magnetic ground states. The strength of $H$ is measured with respect to $H_s$; $H_s=H_r^4/H_{ex}^3$ where $H_{ex}$ and $H_r$ are the exchange and anisotropic fields respectively. In the case of $H<H_s$, there is an existence of correlated spin glass phase which has a large susceptibility and is similar to Ahrony and Pyte phase [23]. Presence of random anisotropy causes variation of the direction of magnetization of locally correlated regimes. In this region, magnetization is expressed as

$$M=M_0\,[1-1/15(H_s/H)^{1/2}]\ldots\ (1)$$

where $M_0$ is the saturation magnetization. In case of high field ($H>H_{ex}$), there is virtual alignment of spins with applied field along with slight tipping angle due to anisotropy. Once the applied magnetic energy overcomes the exchange field energy, $M(H)$ curves saturates. In this region, $M$ is expressed as

$$M=M_0\,(1-1/15(H_r/(H+H_{ex})^2)\ldots.\ (2)$$

The $M$-$H$ data obtained at 2 K for all alloys for $5\leq H\leq 50$ kOe are analyzed using eqn.2 and are shown in Fig. 3 (c). The fit parameters extracted from the fit are given in the table 2. It can be seen that both $H_r$ and $H_{ex}$ increases with Mn concentration. This behaviour reflects an increment in strength of magnetic anisotropy and similar trend has been observed in system exhibiting glassy magnetic phase [13, 24]. In other words, at low temperatures, the strength of anisotropy is large which can result in freezing of correlated spins in glassy state. Hence, it can be said that magnetic anisotropy plays a significant role in the low temperature regime in this series.

**Magnetic phase in low temperature regime**

**Time dependent magnetization study**

To shed some light on the magnetic phase in low temperature regime of this series of alloys isothermal remanent magnetization (IRM) measurements has been performed at different temperatures (Fig. 4) using the following protocol: the alloy is cooled in zero field from room temperature to the measurement temperature (2 and 10 K). After that 100 Oe field is applied for 20 minutes. After that the field is switched off and magnetization as a function of time is recorded. The obtained curves are fitted using equation

$$M(t)= M_0 - S \ln(1+t/t_0)\dots\dots\dots\dots(3)$$

where $M_0$ is the magnitude of magnetization at $t= 0$ and $S$ is the magnetic viscosity [21, 25]. The obtained fitting parameters are tabulated in table 3. For $Fe_2CrAl$, logarithmic relaxation behaviour is observed at 2 K indicating the presence of a large number of intermediate anisotropic metastable states, while this type of behaviour is absent at 10 K (inset of fig. 4(a)). Across the series, similar type of relaxation behaviour is observed at 2 K. It is seen that at 10 K, all Mn doped alloys show relaxation behaviour. It is also observed that magnetic viscosity decreases with increase in temperature possibly due to a variation in the size of the cluster of spins. Usually, for long ranged ordered systems magnetization remains unchanged with time. Our observations indicate the absence of long range ordering and distribution of energy barrier due to the presence of clusters of spins. This indicates a possible presence of a glassy phase in low temperature regime of this series.

**Heat capacity study**

The presence of glassy phase in low temperature regime is also reflected in the detailed analysis of heat capacity measurements. Left panel of Fig. 5 shows the $T^2$ response of $C/T$ for all the alloys. In the low temperature regime the curves are fitted with the following equation

$$C=\gamma T+\beta T^3\dots\dots\dots\dots\dots (4)$$

where $\gamma$ and $\beta$ are the fitting constants. The obtained fitting parameters are listed in table 4. It can be inferred from the table that the obtained values of $\gamma$ for all alloys are quite large. It is also determined that there is increment in the value of $\gamma$ with increasing Mn concentration. Table 4 also contains the value of Debye temperature ($\theta_D$) obtained using $\theta_D= (12\pi^4 nR/5\beta)^{1/3}$ where R is ideal gas constant, n is the number of atoms per formula unit. The values of $\gamma$ and $\beta$ are in accordance with those obtained in case of magnetic glassy systems [26-29]. Also, it can be noted from right panel of Fig. 5 that $C$ deviates from the typical Debye $T^3$ behaviour at low temperature when temperature response of $C/T^3$ is plotted. At low temperature, such upturn indicates the presence of glassy phase and can also be related to the presence of TLS (two level systems) excitation which is in contrast to expected Debye behaviour seen in crystal [26]. The above observations signify the presence of glassy phase in the low temperature regime of $Fe_{2-x}Mn_xCrAl$ (0≤x≤1).

**ZFC and FC memory effect**

In order to discern the origin of low temperature glassy phase in this series of alloys we have performed both FC and ZFC memory effect using stop and wait protocol [30-32]. To investigate the FC memory effect, we have employed the following protocol: the alloy is cooled in the presence of 100 Oe from 300 K to 2 K but cooling is interrupted at different halting temperatures ($T_H$) (for $Fe_2CrAl$, $T_H$=15, 12 and 8 K; for $Fe_{1.75}Mn_{0.25}CrAl$, $T_H$=30 and 15 K; for $Fe_{1.5}Mn_{0.5}CrAl$, $T_H$=10 and 5 K; for $Fe_{1.25}Mn_{0.75}CrAl$, $T_H$= 8 and 4 K and FeMnCrAl $T_H$~8 K) for 2 hours in each case. During the halt, the magnetic field is turned off and after the wait time, it is switched on and cooling is resumed. It is observed that this cooling procedure produces a step-like feature in the obtained FC curve for each sample. This obtained FC magnetization curve is represented as FC-tw in the left panel of Fig. 6. After reaching the 2 K, the alloy is again heated back in presence of 100 Oe to 300 K. Interestingly, the curve obtained after warming also exhibit kinks near the halt temperature. This curve is illustrated as FC curve (in left panel of Fig. 6). It can be said that the alloy remembers its thermal history or magnetic state achieved during FC-tw. Again the alloy is cooled to 2 K without any halt and the warming curve is taken immediately after cooling. The curve is featureless and behaves as normal FC curve which is treated as FC-reference in fig. 6 (left panel). Both SPM and glassy magnetic systems exhibit FC memory effect, but there is a significant difference between the behaviour of FC curves using the above protocol. In the case of SPM, the FC curves increase continuously with decreasing temperature whereas for glassy magnetic systems such behavior is absent [30, 31]. As observed from the figure, there is no increment in magnetization in FC curves with decreasing temperature; implying absence of SPM behaviour in the low temperature region of these alloys. Additionally, according to Ref [30-32] it is reported that ZFC memory effect is exhibited by glassy magnetic system and is not observed in systems exhibiting SPM behaviour. Hence, we have performed ZFC memory effect using the following protocol: the alloy is cooled in the presence of zero field, and halts are made at different temperatures ($T_H$) for different alloys (for $Fe_2CrAl$, $T_H$=7 K, for $Fe_{1.75}Mn_{0.25}CrAl$, $T_H$=5 K, for $Fe_{1.5}Mn_{0.5}CrAl$, $T_H$=5K, for $Fe_{1.25}Mn_{0.75}CrAl$, $T_H$=6 K and for FeMnCrAl, $T_H$=4 K). Then, 100 Oe is applied and data is obtained during the warming cycle (ZFC-tw). The obtained curve is shown in the right panel of Fig. 6. Again the alloy is cooled to 2 K in absence of magnetic field and then 100 Oe is applied and data is taken during warming cycle without any halt. This curve is treated as ZFC-ref. Here, we have observed that there is a difference between $ZFC_{ref}$ and $ZFC_{tw}$ i.e. presence of memory dips at wait temperature (as shown in the inset of the right panel of fig. 6). The observation of ZFC memory effect indicates the presence of a glassy magnetic phase in the low temperature region of this series of alloys. Based on the above observations, it can be said that in the low temperature region the origin of the glassy magnetic phase is due to the presence of interacting spins clusters rather than blocking of spins.

**AC susceptibility study**

In order to understand the nature of the glassy magnetic phase, we have performed frequency dependent AC susceptibility measurements in the absence of any superimposed DC fields. Fig. 7 depicts the temperature response of real ($\chi'$) and imaginary part ($\chi''$) of AC susceptibility for this series, measured at different frequencies (13-931 Hz) for AC drive field of 1 Oe. In the case of $Fe_2CrAl$, it is observed that $\chi'$ exhibits frequency dependent broad feature in low temperature regime, whereas two frequency dependent peaks near $T_{f1}$~12.5 K and $T_{f2}$~3.9 K are clearly visible in $\chi''$. However, $T_{f1}$ is not clearly visible at 13 and 131 Hz. It can be due to the presence of some interacting clusters in this region which display frequency dependent behaviour at higher applied frequencies. In all the Mn doped alloys, $\chi'$ show frequency dependent broad features, while, $\chi''$ exhibit well defined peaks. In $Fe_{1.75}Mn_{0.25}CrAl$, in $\chi''$ two frequency dependent peaks are observed near $T_{f1}$~22 and $T_{f2}$~4.2 K, while, for $Fe_{1.5}Mn_{0.5}CrAl$, three peaks are noted at $T_{f1}$ ~ 30.4, $T_{f2}$ ~ 9.5 K and $T_C$ ~55 K. However, in $Fe_{1.25}Mn_{0.75}CrAl$, two frequency dependent peaks are noted at $T_{f2}$~11.5 K and $T_C$ ~ 27 K. But these peaks near $T_C$ in both compounds does not show any shift with temperature with frequency as compared to $T_{f1}$ and $T_{f2}$. Interestingly, for FeMnCrAl three frequency dependent peaks in $\chi''$ are observed at $T_f^*$ ~22 K , $T_{f1}$~16.6 K and $T_{f2}$~11 K. Hence, it can be said that with increment in Mn concentration, $T_{f1}$ and $T_{f2}$ peaks shift towards high temperature region except in $Fe_{1.25}Mn_{0.75}CrAl$, where only $T_{f2}$ is observed. Presence of $T_f^*$ in FeMnCrAl signifies the absence of long range ordering in the whole temperature range of this alloy.

In all these alloys, it is observed that $T_f^*$, $T_{f1}$, and $T_{f2}$ shift towards higher temperature as the magnitude of measuring frequency increases. Such a frequency dependent shift can be due to either glass-like freezing or SPM blocking [21]. However, the presence of SPM-like behaviour in these systems has been ruled out in the earlier sections. Thus, in order to characterize the nature of glassy transition exhibited by $T_f^*$, $T_{f1}$, and $T_{f2}$ in all alloys we have determined Mydosh parameter which is often used to distinguish/compare various magnetic glassy systems. This parameter reflects the response of magnetic clusters toward frequencies and it depends on the type of interaction between clusters. As the interaction between magnetic clusters is generally weak, it results in strong sensitivity towards frequencies. In contrast for normal magnetic systems, the interactions between the magnetic ions are strong and such frequency dependence is not observed. Mydosh parameter ($\delta T_f$) is expressed as [21]

$$\delta T_f = \Delta T_f / T_f \log f \ldots\ldots\ldots\ldots\ldots (5)$$

where $T_f$ is the peak temperature corresponding to frequency ($f$). The obtained Mydosh parameters are listed in table 5. For $Fe_2CrAl$, it is found that the value of $\delta T_f$ for $T_{f2}$ is similar to that reported for other CG systems [21, 33-37], whereas it is not calculated for $T_{f1}$. For the Mn-doped alloys, the values of $\delta T_f$ obtained for the respective freezing temperatures indicate that they also belong to the CG category. We have further investigated the behaviour of magnetic clusters, by analyzing the temperature dependence of relaxation time ($\tau$) using standard critical slowing down model, which is given by dynamic scaling theory [21, 38]

$$\tau=\tau^* \left((T_f-T_g)/(T_g)\right)^{-z\nu} \ldots.. (6)$$

where $\tau^*$ is the microscopic flipping time, $T_g$ is the true spin glass (SG) transition temperature, $z$ is the dynamic critical exponent and $\nu$ is the critical exponent of the correlation length. Also, the temperature maxima is fitted with Vogel-Fulcher (V-F) law (which takes into account the interaction between spins) of the form [21,39]

$$\tau=\tau_0 \exp\left((E_a)/k_B(T_f-T_0)\right) \ldots\ldots (7)$$

where $T_0$ is the VF temperature, which represents the strength of interactions between clusters. The fittings using equation (6) and (7) is shown in Fig. 8. The values of parameters obtained after fitting are summarized in table 6. In the case of $Fe_2CrAl$, we have obtained non-zero value of $T_0$ and our data agrees well with V-F law which suggests a finite interaction among spins and thus leading to formation of clusters [40]. It can also be inferred that the magnitude of $\tau^*$ and $\tau_0$ obtained in case of $T_{f2}$ is quite higher than that obtained in case of SG systems [$10^{-10}$-$10^{-13}$ s]. Such large values have been observed in various CG systems [21, 25]. For all the Mn-doped alloys, for each freezing temperature, the value of $T_g$ is found large than $T_0$ which is in accordance with the trend found in CG systems. Similarly, the obtained values of $z\nu$ for all the freezing temperature also support the preceding statement [41-43]. Thus, it can be said that $T_f^*$, $T_{f1}$, and $T_{f2}$ correspond to CG-type freezing in the respective alloys. It also suggests that this series of alloys exhibit slow spin dynamics due to cluster formation. This type of existence of more than one glassy transition in a system is not unusual; it has been reported in various systems [41, 42].

In order to see the effect of anisotropic behaviour in cluster glass region, we have also performed AC susceptibility measurements in the presence of superimposed DC fields (0-1500 Oe) for all alloys by fixing the AC excitation field at 1 Oe and 531 Hz. Field response of AC susceptibility is plotted in Fig. 9. These measurements will help to discern the highly anisotropic from less anisotropic cluster region. In the presence of large DC fields, the less anisotropic regions get aligned and their responses to small AC fields are suppressed [44]. As it can be seen from Fig. 9, in case of $Fe_2CrAl$, $T_{f2}$ transition shifts towards lower temperature with increasing $H_{DC}$, thereby indicating the highly anisotropic nature of CG transition. A different behaviour is observed in the case of $T_{f1}$. In the case of $T_{f1}$, the peak gets transformed into a broad shoulder when $H_{DC}$ is applied and gets suppressed under the application of 750 Oe. Similar shifting of $T_{f2}$ peak towards lower temperature and transformation of $T_{f1}$ peak into broad shoulder along with suppression in the presence of 500 and 300 Oe is observed in case of $Fe_{1.75}Mn_{0.25}CrAl$ and $Fe_{1.5}Mn_{0.5}CrAl$ respectively. For $Fe_{1.25}Mn_{0.75}CrAl$, it is observed that $T_{f2}$ shifts towards lower temperature. However, in case of FeMnCrAl, it is found that $T_f^*$ and $T_{f1}$ combines into single broad shoulder like feature which suppresses under 300 Oe. This type of behaviour exhibited by $T_{f1}$ in all alloys (except $Fe_{1.25}Mn_{0.75}CrAl$) can be due to presence of less anisotropic cluster regions near $T_{f1}$ which gets aligned in the direction of applied $H_{DC}$ and their response to AC field gets suppressed. It is also noted that on the application of the DC field, $T_{f2}$ moves towards lower temperature. Hence, we have focussed on the response of $T_{f2}$ towards applied $H_{DC}$. In the case of glassy magnetic systems, typically two irreversible lines are observed in *H*-*T* phase diagram: de Almeida-Thouless line (*A*-*T* line) and Gabay-Toulose (*G*-*T* line) line. In case of Heisenberg spin systems, we can observe both the lines, in the strong anisotropy regime we can expect that the line follows *A*-*T* character while in weak anisotropic region G-T behaviour is followed. On other hand, only *A*-*T* line is usually observed for Ising spin glass systems [45, 46]. However, we can expect a quantitative difference in *A*-*T* line in the mean field and non mean field model. According to non-mean field theory, in low field region the variation of $T_f$ with $H_{DC}$ follows [47]

$$T_f\,(H) = T_f\,(0)(1-A'H^Q)\ldots\ldots\ldots\ldots(8)$$

where $A'$ is the anisotropic parameter, $T_f$ (0) is the value of freezing temperature in the absence of magnetic field and $Q$ is the exponent. In the mean field model, $Q$ has a value of 0.66 in strong anisotropic regime whereas it has a value of 2 in the weak anisotropic regime. The obtained parameters from fitting the response of $T_{f2}$ on the application of $H_{DC}$ (with eqn.

8) are mentioned in table 7. The obtained values of $Q$ corresponding to each alloy reflect that $T_{f2}$ does not follow the mean field theory; rather a non-mean field type exponent has been obtained similar to other SG systems which do not belong to same universality class [48, 49]. Also, from the table, it can be inferred that all alloys neither follows *A-T* line nor *G-T* line but it can be said that as Mn concentration increases, the system transits from weak irreversibility to strong irreversibility regime (Fig. 10). Therefore, it can be said that Mn concentration increases the role of anisotropy in the system. Also, we would like to mention that, in case of $Fe_2CrAl$ it is observed that $\chi''$ near $T_C$ shows frequency dependent behaviour i.e. there is a decrement in magnitude with increment in frequency (not shown) but no shifting of peak temperature with frequency is noted. This type of frequency dependence can be due to irreversible domain wall movement or pinning effect [50]. Similar type of frequency dependent behaviour is also observed in $Fe_{1.75}Mn_{0.25}CrAl$, $Fe_{1.5}Mn_{0.5}CrAl$ and $Fe_{1.25}Mn_{0.75}CrAl$. Hence, it implies that magnetic anisotropy plays an important role and affects the nature of magnetic phase for the various concentration of Mn. This is also responsible for the absence of long range magnetic ordering in the FeMnCrAl compound.

**Evidence of the presence of the Griffiths phase in the high temperature region**

As mentioned in section "Temperature and magnetic field dependent DC magnetization studies", in PM phase of $Fe_{1.5}Mn_{0.5}CrAl$, a weak feature is noted. In order to see whether the similar feature is present in $Fe_2CrAl$ and $Fe_{1.75}Mn_{0.25}CrAl$, the temperature response of magnetization is measured upto 400 K at 100 Oe (not shown). It is observed that there is a weak bifurcation between ZFC and FC curves in high temperature regime, which indicates to the presence of short range correlations among spins. Additionally, similar to $Fe_{1.5}Mn_{0.5}CrAl$, some anomaly is noted around 364 and 300 K for $Fe_2CrAl$ and $Fe_{1.75}Mn_{0.25}CrAl$ respectively. To shed some light on the observed anomaly in this series of alloy, we have plotted temperature response of inverse DC susceptibility ($\chi^{-1}_{DC}$) (left panel of Fig. 11). For all these alloys, a strong downward deviation from Curie Weiss (CW) law is observed. For $Fe_2CrAl$ it is observed near 364 K and is shifted to 300, 206, 180 and 214 K for $Fe_{1.75}Mn_{0.25}CrAl$, $Fe_{1.5}Mn_{0.5}CrAl$, $Fe_{1.25}Mn_{0.75}CrAl$ and FeMnCrAl respectively. It has been reported in the literature that the observation of a downward deviation of $\chi^{-1}_{DC}$ curve well above $T_C$ is a signature of the presence of the GP regime [51, 52]. Hence the observation of similar feature gives an indication of the possibility of the presence of GP regime in this series of alloys. Under the application of the magnetic field, it is noticed that the downturn softens (shown in figure 11 (k) and (l) corresponding to $Fe_2CrAl$ and FeMnCrAl, respectively). Similar suppression of the downturn has also been reported in the GP regime of other magnetic systems [53-55]. In general, Griffith's singularity can be characterized using the exponent ($\lambda$) which can be obtained from the expression

$$\chi^{-1} \propto (T-T_C^R)^{(1-\lambda)} \ldots\ldots (9)$$

where $0 \leq \lambda < 1$ and $T_C^R$ is the critical $T$ where $\chi$ diverges. This relation is the modified form of CW law where $\lambda$ signifies the deviation from CW behaviour. This value has a finite value between 0 and 1 above $T_C$ but is approximately 0 in PM region. We have plotted $\log_{10} \chi^{-1}$ vs $\log_{10} (T/T_C^R-1)$ (right panel of fig. 11) and the value of $\lambda$ is obtained from the slopes, fitted with eqn. 9. In PM and GP region, the exponent is defined as $\lambda_P$ and $\lambda_G$ respectively. The obtained values are mentioned in table 8. For all these alloys, we have observed larger values of $\lambda_G$ (with the value lying between 0 and 1) which signifies the strong deviation from CW behaviour in high temperature region. It is also observed that as Mn concentration is increased, the GP becomes stronger. This behaviour can be attributed to the formation of short range clusters. Such features are in accordance to the theoretical studies, where it is argued that coexistence of competing magnetic phases stabilizes as well as enhances the possibility of clusters to be present above $T_C$, resulting in an inhomogeneous PM phase [56]. Interestingly, AC susceptibility also shows an anomaly in this high temperature region. However, no shift in maxima with frequency in AC susceptibility is observed in all alloys but it suppresses under the application of superimposed $H_{DC}$=100 Oe (not shown), which is in contrast to that observed for the clusters in the low temperature regime. These observations rule out the possibility of formation of glassy magnetic state or SPM-like behaviour in this temperature region. In low temperature regime, anisotropy is dominant and plays a crucial role in the formation of CG phase but the origin of clusters in this temperature regime is different from those in low temperature regime. Here, they are formed as a result of anti-site disorder between Fe and Al [57, 58, and 59] and anisotropy has negligible effect in this regime. It is interesting to note that $Fe_2CrAl$ also shows the presence of this phase inspite of its stoichiometric composition. As we have noted from figure 1(b) and inset of (c) that the intensity ratio of $I_{200}/I_{220}$ is low (2.7%), which indicates the presence of substantial amount of anti-site disorder between Fe/Mn and Al. This anti-site disorder is responsible for the observation of GP in the parent compound. It is also observed that the disorder increases with Mn concentration (as discussed in section "Structural analysis"). A similar trend is noted in the value of $\lambda_G$, which reflects the role of anti-site disorder in the formation of GP in these alloys. It results in the formation of the highly inhomogeneous phase where different values of exchange coupling are allocated randomly at different lattice sites. It may lead to coexistence of short range FM and AFM correlations. Such anti-site disorder and competition between AFM-FM interactions are responsible for evolution of GP in these alloys.

**Discussions**

The Heusler alloy $Fe_2CrAl$ crystallizes in $L2_1$ structure. The Mn substituted $Fe_2CrAl$ alloys have similar XRD spectra but with increasing Mn concentration, significant changes in the intensity of (200) reflection line and increment in lattice parameter are noted. These observed changes can be due to atomic site disorder between Fe and Al arising out of substitution of Mn in crystal lattice, which also plays a significant role in altering the magnetic correlations in the system. The observed drastic decrease in value of $T_C$ with increment in Mn concentration is due to development of AFM coupling between Fe and Mn. Presence of more than one CG phase transition is due to the fact that magnetic anisotropy plays a significant role and it increases with the increment in Mn concentration. At low temperature, this anisotropy grows larger such that it weakens the coupling that holds together infinite clusters. Hence, these clusters dissociate into smaller clusters. It is noted that less anisotropic and bigger clusters freeze at higher temperature (near $T_f^*$ and $T_{f1}$); whereas those with higher anisotropic field freezes at lower temperature (near $T_{f2}$). Along with anisotropy, site disorder also plays a significant role in altering the magnetic properties of this series of alloys. GP-like features are observed high temperature regime due to the presence of anti-site disorder, which increases with Mn concentration.

**Phase diagram of $Fe_{2-x}Mn_xCrAl$ ($0 \leq x \leq 1$):**

Figure 12 depicts the *T-x* phase diagram of $Fe_{2-x}Mn_xCrAl$. The phase boundaries in the diagram are estimated from the DC magnetization and AC susceptibility studies. Phase (CG-I) is the CG phase it is composed of small sized and highly anisotropic clusters. At higher temperature, there is another CG phase (Phase CG-II) consisting of less anisotropic and bigger sized clusters. However, this phase is absent in $Fe_2CrAl$ and $Fe_{1.25}Mn_{0.75}CrAl$. Phase (CG-III) (present only in FeMnCrAl) also has similar character as phase CG-II. Phase FM represents the FM phase of this series. This phase gets suppressed with increment in Mn due to development of AFM coupling between Fe and Mn and it vanishes in FeMnCrAl. The intermediate phase (IM) is the region lying between the FM and GP phase region. In this phase, as observed from magnetization and AC susceptibility results, there is an abrupt decrease in magnetization value as the temperature is increased (after $T_C$). Such abrupt drop has been noted in ferromagnetic systems, when phase transition takes place from ordered (FM) to disordered phase. In high temperature region, GP (Phase GP) is observed, which gets suppressed towards lower temperature with increment in Mn content. Phase PM is the PM phase of these alloys.

## Summary

In summary, it is concluded that $Fe_2CrAl$ is a ferromagnetic Heusler alloy crystallizing in $L2_1$ structure with $T_C$~202 K and exhibits CG state in low temperature regime. With increase in Mn concentration, $T_C$ is significantly suppressed towards lower temperature. However, it is noted that below $T_C$, $Fe_{1.75}Mn_{0.25}CrAl$ and $Fe_{1.5}Mn_{0.5}CrAl$ exhibit double CG-like transitions. Interestingly for $Fe_{1.25}Mn_{0.75}CrAl$; a single CG-like transition is noted below $T_C$. For FeMnCrAl, no long ranged magnetic ordering is observed and this alloy undergoes three CG-like transitions. In this series of alloys, observation of more than one CG state has been attributed to increment in magnetic anisotropy in the system. It weakens the coupling between infinite clusters and dissociates them into small clusters. These clusters freeze at different temperatures depending upon their anisotropic fields. In present case, we have found that bigger and less anisotropic clusters are near $T_f^*$ and $T_{f1}$ whereas smaller and highly anisotropic near $T_{f2}$. Additionally, at high temperatures, GP phase is observed in $Fe_2CrAl$ which is shifted towards lower temperature with increasing Mn concentration and this feature arises as a result of anti-site disorder between Fe and Al.

## Methods

The Heusler alloys $Fe_{2-x}Mn_xCrAl$ (x= 0, 0.25, 0.5, 0.75 and 1) are prepared by arc melting the stoichiometric ratio of the high purity of the constituent elements (>99.9%) in an atmosphere of argon. The ingots are re-melted several times to ensure the homogeneity of the alloys. The weight loss after the final melting for each alloy is less than 1%. The resultant ingots are sealed in evacuated quartz tubes and subjected to 900 $^0$C for 1 week, followed by water quenching. The elemental analysis is carried using energy dispersive x-ray analyzer attached with FESEM (Nova Nano SEM-450, JFEI U.S.A). The average atomic stoichiometry of each alloy is in accordance with the expected values. X-ray diffraction (XRD) measurements are performed at room temperature at P02 beamline in the Petra III Synchrotron radiation, Hamburg, Germany using a wavelength of 0.20712 Å. Temperature ($T$) and magnetic field ($H$) dependent magnetization ($M$) measurements are carried using Magnetic property measurement system (MPMS) from Quantum Design, U.S.A. Heat capacity (*C*) is measured using the heat capacity option of Physical property measurement system (PPMS) from Quantum design, U.S.A.

**Acknowledgments**

The authors acknowledge IIT Mandi for providing the experimental facilities. KM acknowledges the financial support from a research grant (grant no: 03(1381)/16/EMR-II) from SERB, India. The support from DST under its India-DESY scheme through Jawaharlal Nehru Centre for Advanced Scientific Research is acknowledged. SS thanks Science and Engineering Research Board of India for financial support through the award of Ramanujan Fellowship (grant no: SB/S2IRJN-015/2017) and Early Career Research Award (grant no: ECR/2017/003186).

**Author Contributions**

KY and MKS synthesized the alloys. KY performed the experiments. SS performed the synchrotron X-ray diffraction experiments and its analysis. KY and KM analysed the data and wrote the manuscript in consultation with other authors.

## Additional Information

The authors declare no competing interests

**Tables-**

Table 1: Lattice parameters of $Fe_{2-x}Mn_xCrAl$ ($0 \le x \le 1$) obtained from Le-bail fit of XRD patterns.

| | Lattice parameter (Å) | Unit cell volume (Å$^3$) |
|---|---|---|
| $Fe_2CrAl$ | 5.784±0.0002 | 193.383±0.0008 |
| $Fe_{1.75}Mn_{0.25}CrAl$ | 5.791±0.0002 | 194.285±0.0008 |
| $Fe_{1.5}Mn_{0.5}CrAl$ | 5.800±0.0001 | 195.112±0.0001 |
| $Fe_{1.25}Mn_{0.75}CrAl$ | 5.804±0.0002 | 195.536±0.0008 |
| FeMnCrAl | 5.815±0.0001 | 196.629±0.0001 |

Table 2: Parameters obtained from RAT fitting using eqn. 2 of *M-H* data at 2 K

| | $M_0$ (emu/gm) | $H_{ex}$ (kOe) | $H_r$ (kOe) |
|---|---|---|---|
| $Fe_2CrAl$ | 46.3±0.07 | 29.1±1.9 | 29.0±1.8 |
| $Fe_{1.75}Mn_{0.25}CrAl$ | 39.3±0.11 | 40.9±2.4 | 62.6±2.8 |
| $Fe_{1.5}Mn_{0.5}CrAl$ | 26.2±0.2 | 42.7±2.4 | 101.9±4.9 |
| $Fe_{1.25}Mn_{0.75}CrAl$ | 20.7±0.2 | 45.8±2.2 | 117.0±8.3 |
| FeMnCrAl | 14.8±0.2 | 52.9±3.9 | 156.1±11.3 |

Table 3: Parameters obtained from fitting of time dependent magnetization data with eqn. 3

| | T (K) | $M_0$ (emu/gm) | $S$ (emu/gm) |
|---|---|---|---|
| $Fe_2CrAl$ | 2 | 0.573±0.0004 | 0.002±0.000002 |
| $Fe_{1.75}Mn_{0.25}CrAl$ | 2 | 4.59±0.001 | 0.138±0.0001 |
| | 10 | 0.33±0.0004 | 0.01±0.00002 |
| $Fe_{1.5}Mn_{0.5}CrAl$ | 2 | 7.89±0.001 | 0.234±0.0001 |
| | 10 | 0.42±0.0007 | 0.01±0.00004 |
| $Fe_{1.25}Mn_{0.75}CrAl$ | 2 | 4.30±0.001 | 0.129±0.0001 |
| | 10 | 0.28±0.0004 | 0.011±0.00002 |
| FeMnCrAl | 2 | 0.29±0.0003 | 0.008±0.00002 |
| | 10 | 0.15±0.0002 | 0.002±0.00002 |

Table 4: Parameters obtained from fitting of *C/T* Vs. $T^2$ using eqn. 4

| | $\gamma$ (mJ/mole-$K^2$) | $\beta$ (mJ/mole-$K^4$) | $\theta_D$ (K) |
|---|---|---|---|
| $Fe_2CrAl$ | 38.2±0.09 | 1.1±0.02 | 191.7 |
| $Fe_{1.75}Mn_{0.25}CrAl$ | 47.6±0.2 | 0.32±0.007 | 289.4 |
| $Fe_{1.5}Mn_{0.5}CrAl$ | 55.1±0.007 | 0.43±0.003 | 262.2 |
| $Fe_{1.25}Mn_{0.75}CrAl$ | 59.4±0.01 | 0.44±0.002 | 260.2 |
| FeMnCrAl | 61.4±0.01 | 0.22±0.001 | 329.0 |

Table 5: Mydosh parameters obtained from eqn. 5.

| | $\delta T_f^*$ | $\delta T_{f1}$ | $\delta T_{f2}$ |
|---|---|---|---|
| $Fe_2CrAl$ | - | | 0.18 |
| $Fe_{1.75}Mn_{0.25}CrAl$ | - | 0.06 | 0.17 |
| $Fe_{1.5}Mn_{0.5}CrAl$ | - | 0.08 | 0.11 |
| $Fe_{1.25}Mn_{0.75}CrAl$ | - | - | 0.16 |
| FeMnCrAl | 0.06 | 0.02 | 0.03 |

Table 6: Parameters obtained from critical slowing down model and V-F law using eqn. 6 and 7

| | | $T_g$(K) | $z\nu$ | $\tau^*$ (s) | $\tau_0$ (s) | $E_a/k_B$ (K) | $T_0$ (K) |
|---|---|---|---|---|---|---|---|
| $Fe_2CrAl$ | $T_{f2}$ | 3.5 | 3.4±0.2 | 6.3X10$^{-6}$ | 3.2X10$^{-6}$ | 7.29±0.23 | 3 |
| $Fe_{1.75}Mn_{0.25}CrAl$ | $T_{f1}$ | 21.5 | 2.7±0.3 | 3.2X10$^{-7}$ | 1.5X10$^{-5}$ | 6.61±0.43 | 21 |
| | $T_{f2}$ | 3.25 | 6.1±0.4 | 5.8X10$^{-7}$ | 8.5X10$^{-7}$ | 11.43±0.65 | 3 |
| $Fe_{1.5}Mn_{0.5}CrAl$ | $T_{f1}$ | 30 | 1.8±0.1 | 4X10$^{-6}$ | 3.3X10$^{-5}$ | 7.97±0.17 | 29 |
| | $T_{f2}$ | 9 | 3.1±0.2 | 2X10$^{-7}$ | 2.9X10$^{-6}$ | 12.67±0.92 | 8 |
| $Fe_{1.25}Mn_{0.75}CrAl$ | $T_{f2}$ | 11 | 2.4±0.1 | 9.8X10$^{-7}$ | 1.4X10$^{-5}$ | 10.15±0.29 | 10 |
| FeMnCrAl | $T_f^*$ | 21.75 | 2.4±0.3 | 2.3X10$^{-7}$ | 2.8X10$^{-5}$ | 3.01±0.19 | 21.5 |
| | $T_{f1}$ | 16.2 | 5±0.2 | 1.01X10$^{-10}$ | 2.3X10$^{-6}$ | 5.41±0.09 | 16 |
| | $T_{f2}$ | 10.7 | 2.8±0.4 | 3.3X10$^{-7}$ | 3.8X10$^{-5}$ | 3.01±0.42 | 10.7 |

Table 7: Parameters obtained from fitting of field response of freezing temperature using eqn.8

| | $T_f(0)$ (K) | A | Q |
|---|---|---|---|
| $Fe_2CrAl$ | 5.17±0.27 | 0.57±0.20 $kOe^{-1.30}$ | 1.30±0.69 |
| $Fe_{1.75}Mn_{0.25}CrAl$ | 4.98±0.02 | 0.28±0.03 $kOe^{-1.28}$ | 1.28±0.18 |
| $Fe_{1.5}Mn_{0.5}CrAl$ | 10.74±0.23 | 0.88±0.06 $kOe^{-1.04}$ | 1.04±0.12 |
| $Fe_{1.25}Mn_{0.75}CrAl$ | 13.04±0.36 | 0.77±0.05$kOe^{-0.46}$ | 0.46±0.06 |
| FeMnCrAl | 12.00±0.18 | 0.51±0.1 $kOe^{-0.29}$ | 0.29±0.02 |

Table 8: Parameters obtained from straight line fitting of log-log plot of $\chi_{DC}^{-1}$ as function of $(T/T_C^R)$-1 in PM and GP region

| | $T_C^R$(K) | $\lambda_P$ | $\lambda_{GP}$ |
|---|---|---|---|
| $Fe_2CrAl$ | 313 | 0.0004 | 0.84±0.02 |
| $Fe_{1.75}Mn_{0.25}CrAl$ | 244.5 | 0.0075 | 0.87±0.05 |
| $Fe_{1.5}Mn_{0.5}CrAl$ | 148 | 0.008 | 0.94±0.01 |
| $Fe_{1.25}Mn_{0.75}CrAl$ | 110 | 0.013 | 0.95±0.02 |
| FeMnCrAl | 139 | 0.003 | 0.97±0.01 |

**Figures**

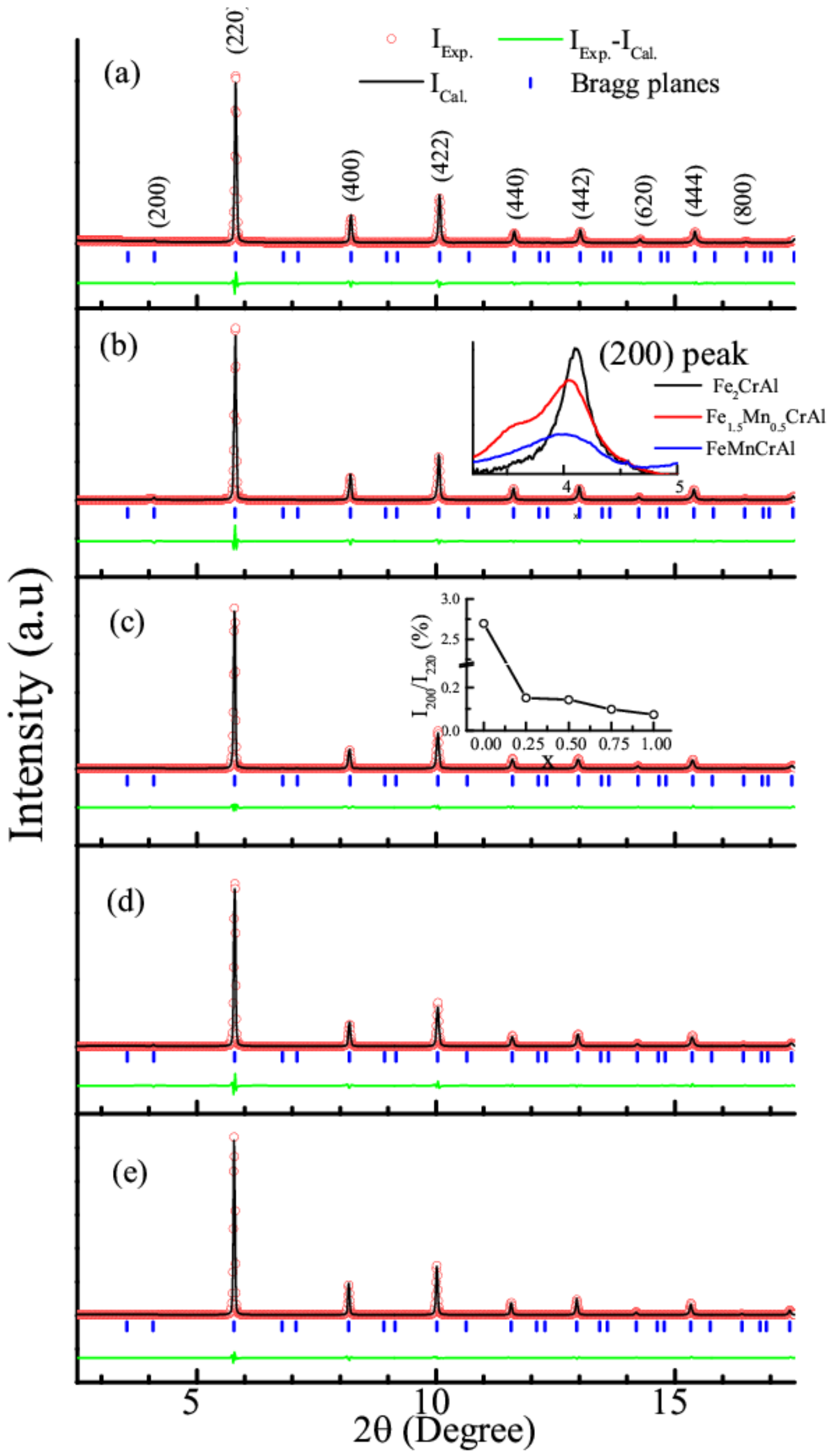

Figure 1 Le-bail fit of the room temperature indexed XRD patterns of (a) $Fe_2CrAl$ (b) $Fe_{1.75}Mn_{0.25}CrAl$ (c) $Fe_{1.5}Mn_{0.5}CrAl$ (d) $Fe_{1.25}Mn_{0.75}CrAl$ and (e) FeMnCrAl. Inset of (b) shows the shifting of peak with increasing Mn substitution along with decrease in intensity of (200) peak for selected compositions of $Fe_2CrAl$, $Fe_{1.5}Mn_{0.5}CrAl$ and FeMnCrAl. Inset of (c) depicts the change in intensity ratio of $I_{200}/I_{220}$ reflection lines with Mn substitution.

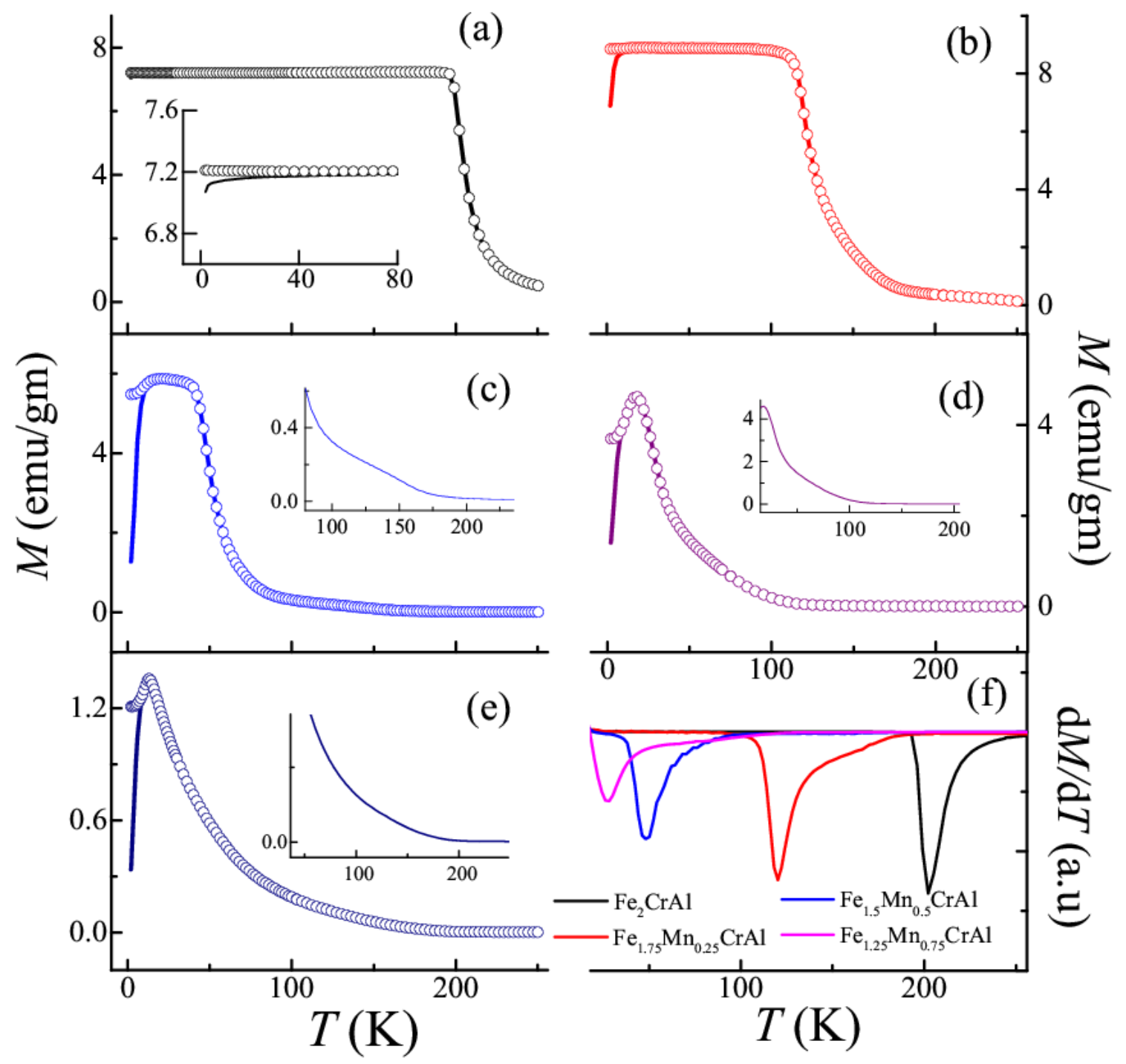

Figure 2 Temperature response of the DC magnetization under ZFC (straight line) and FC (open circle) conditions at 100 Oe in the temperature range 2-250 K for (a) $Fe_2CrAl$. (b) $Fe_{1.75}Mn_{0.25}CrAl$ (c) $Fe_{1.5}Mn_{0.5}CrAl$; (d) $Fe_{1.25}Mn_{0.75}CrAl$; (e) FeMnCrAl. Inset of (a) shows the magnified view of ZFC and FC curves in the temperature range 2-80 K, while, insets of (c-e) shows magnified view of the respective ZFC curves (f) d*M*/d*T* vs *T* plot for $Fe_{2-x}Mn_xCrAl$ (x≤0<1) alloys in the temperature range 18-250 K.

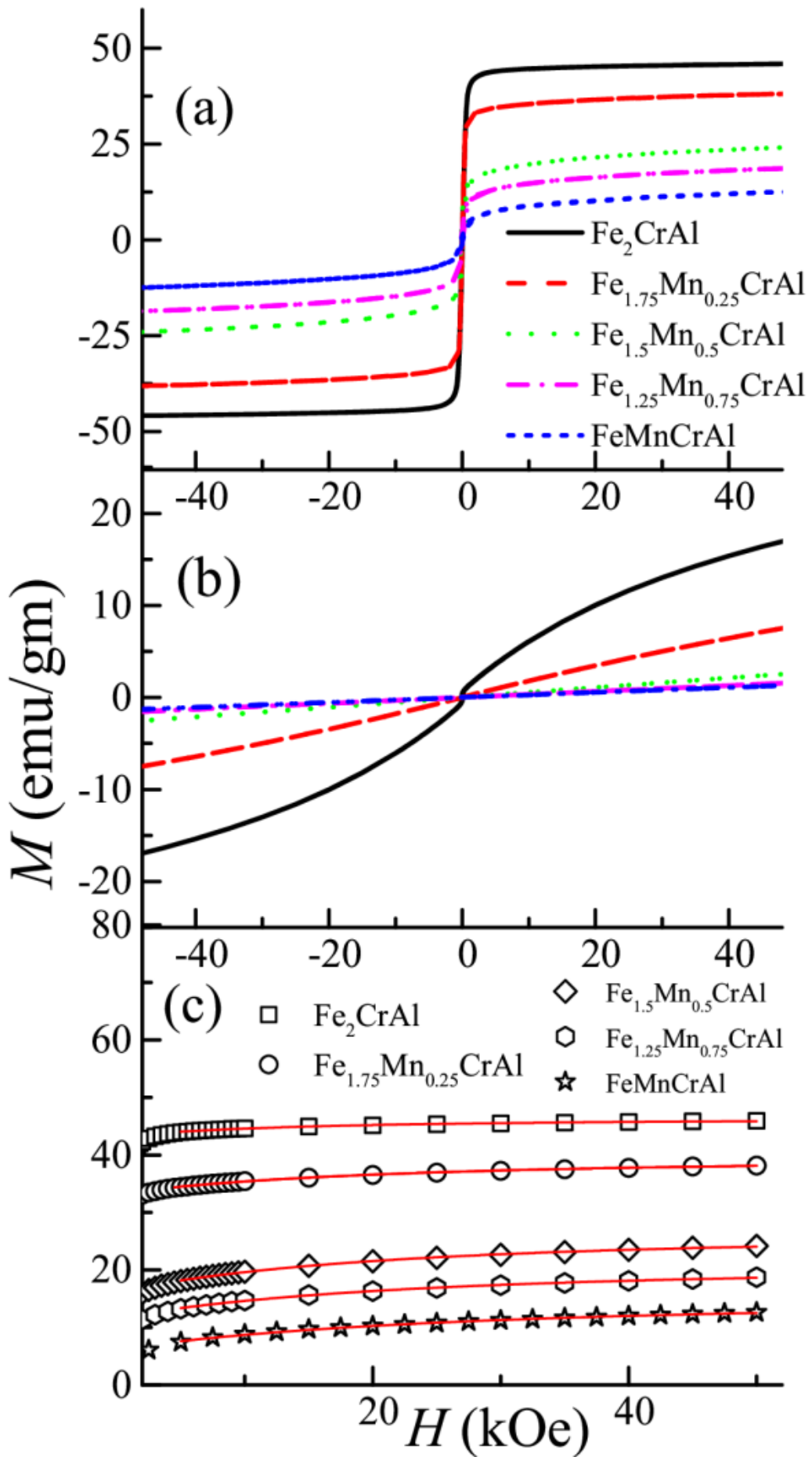

Figure 3 Isothermal magnetization as a function of magnetic field at (a) 2 K and (b) 300 K of $Fe_{2-x}Mn_xCrAl$ (x≤0≤1) alloys (c) Magnetic field dependent magnetization plots of all alloys at 2 K with fit (red solid line) using eqn. 2.

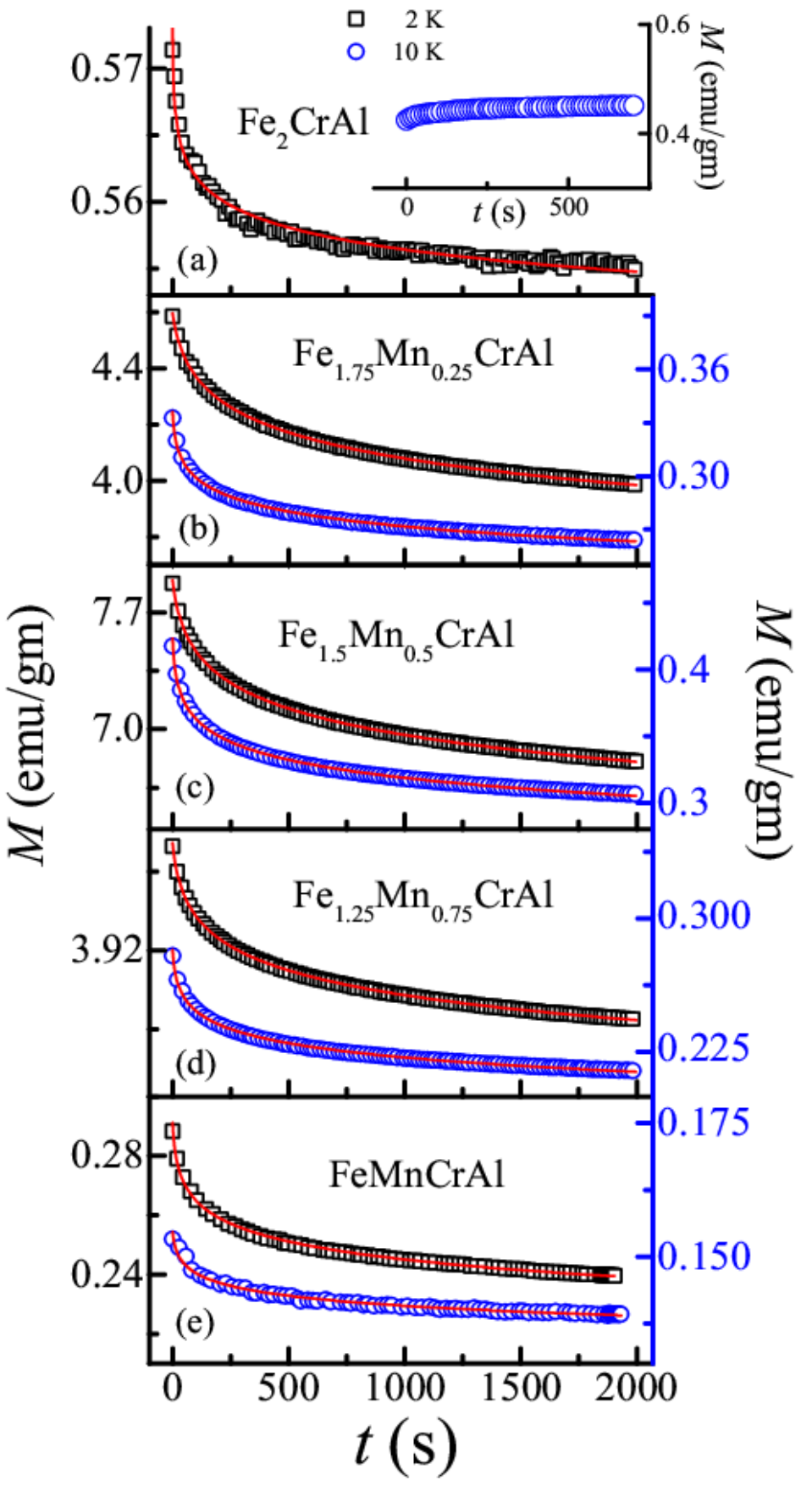

Figure 4 (a), (b), (c), (d) and (e) Magnetization as a function of time at different temperature for all the alloys. Inset of (a) represents the relaxation behaviour of $Fe_2CrAl$ at 10 K. Solid red line depicts the logarithmic fit using eqn. 3. Note: Left axis represents the scale for data obtained at 2 K and right axis represents the scale for the data obtained at 10 K.

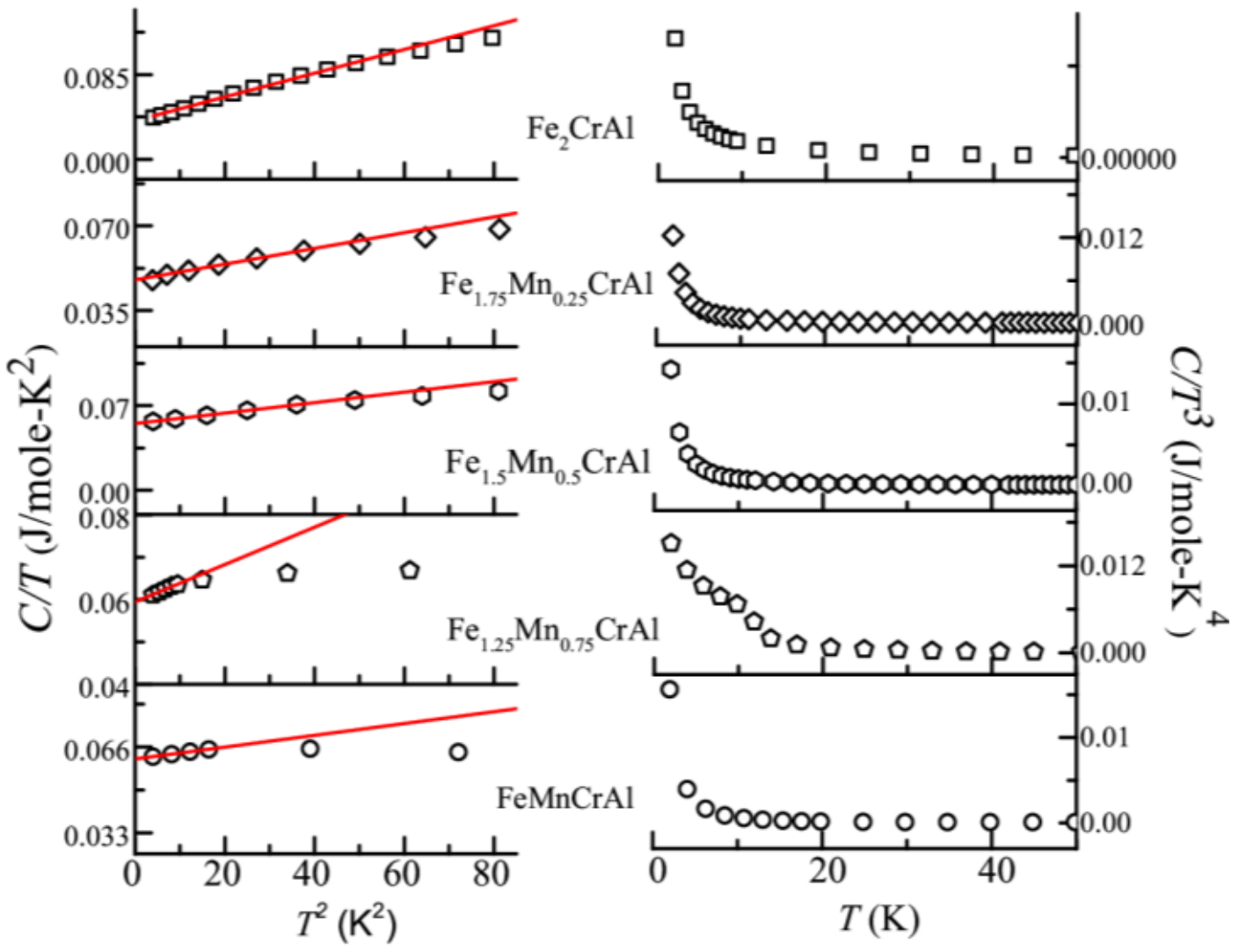

Figure 5 Left panel: $C/T$ vs. $T^2$ plot of all the alloys and fitted (solid red line) with eqn. 4. Right panel: $C/T^3$ vs. $T$ plot of all the alloys.

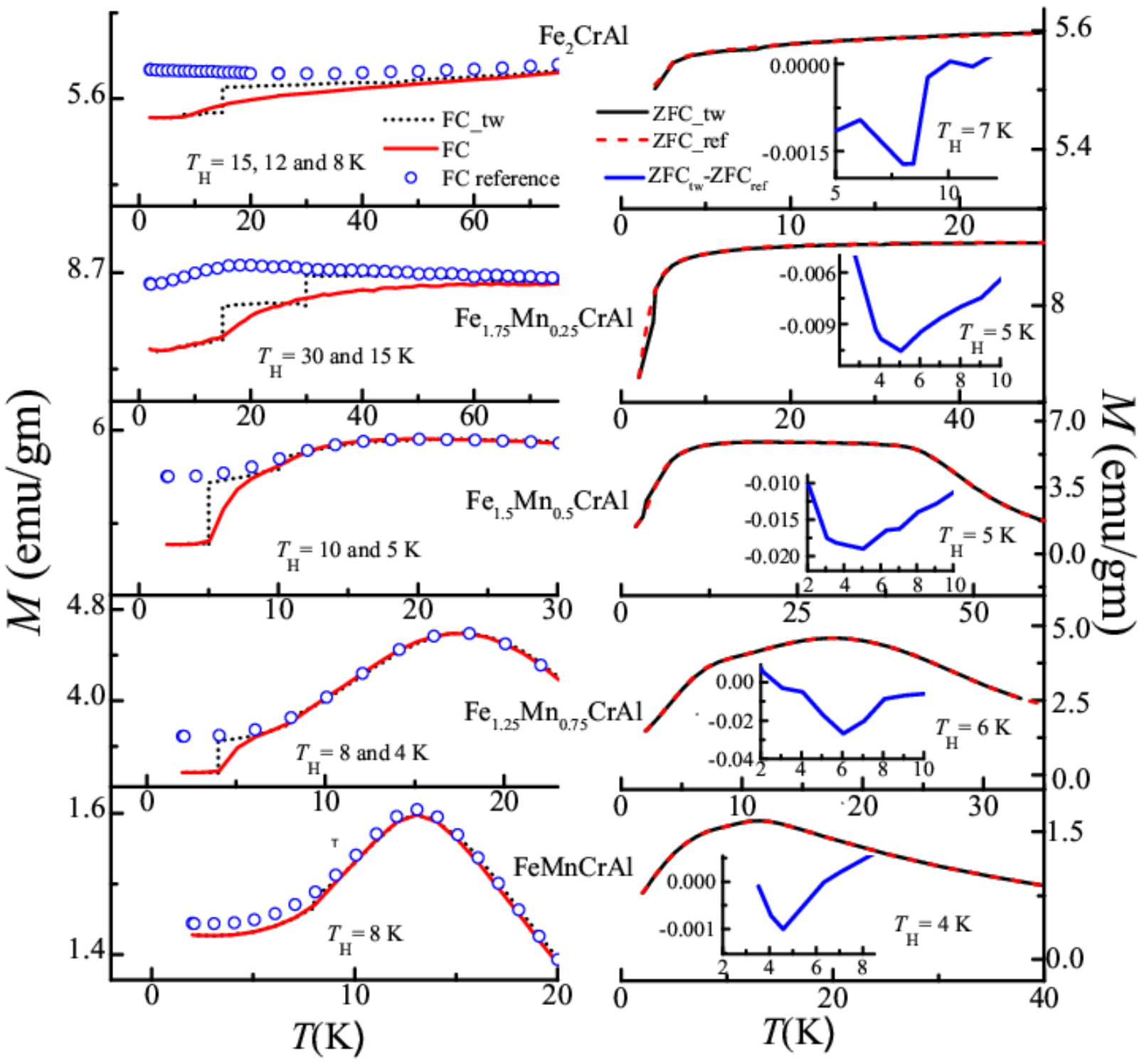

Figure 6 Temperature dependent memory effect of all the alloys at different halt temperatures ($T_H$); Left panel: Under FC condition. Right panel: Under ZFC condition. Insets of right panel: Temperature response of $ZFC_{tw}$-$ZFC_{ref}$.

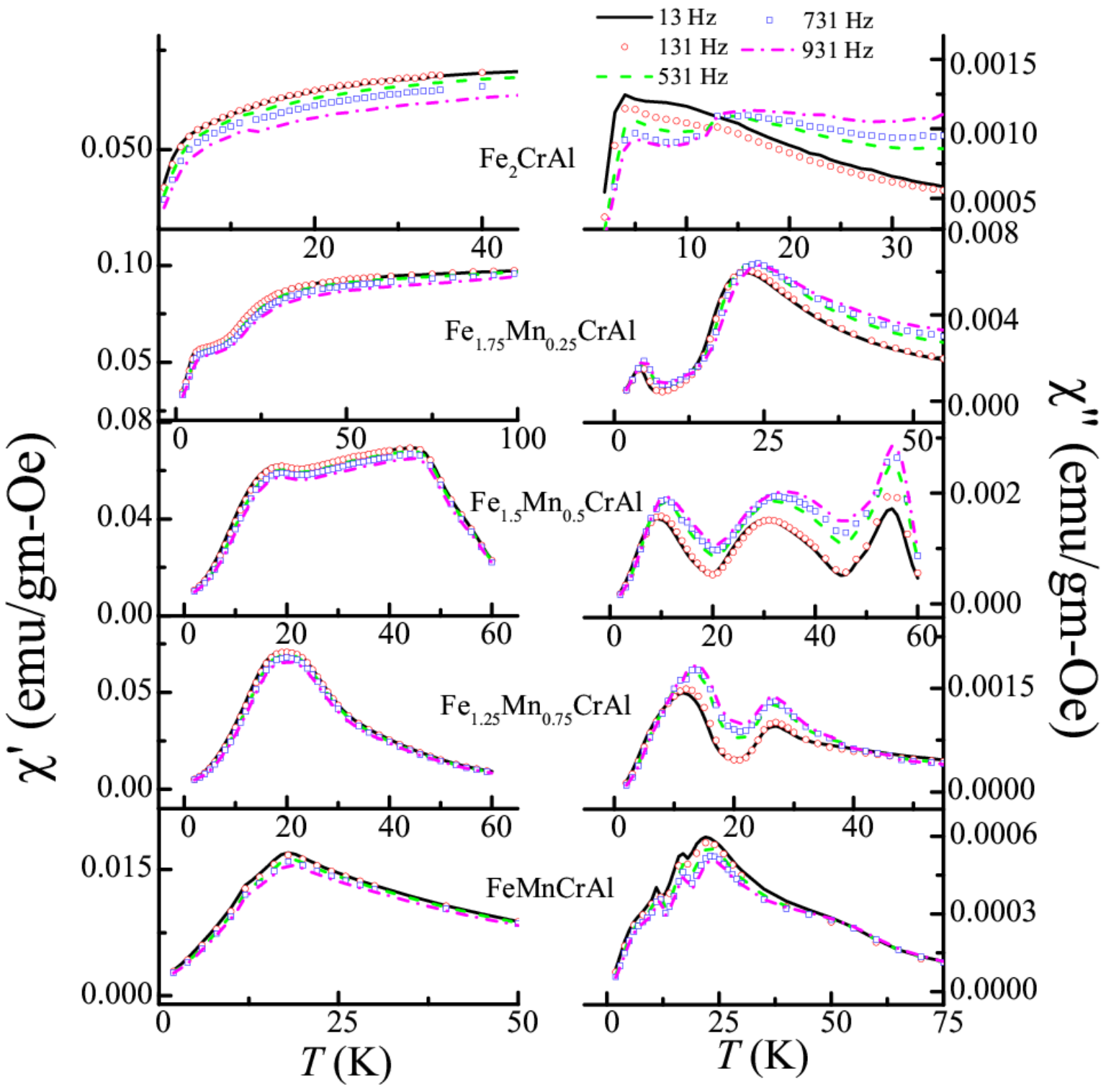

Figure 7 Temperature response of in-phase (left panel) and out-of-phase (right panel) components of AC susceptibility of all the alloys measured at different frequencies.

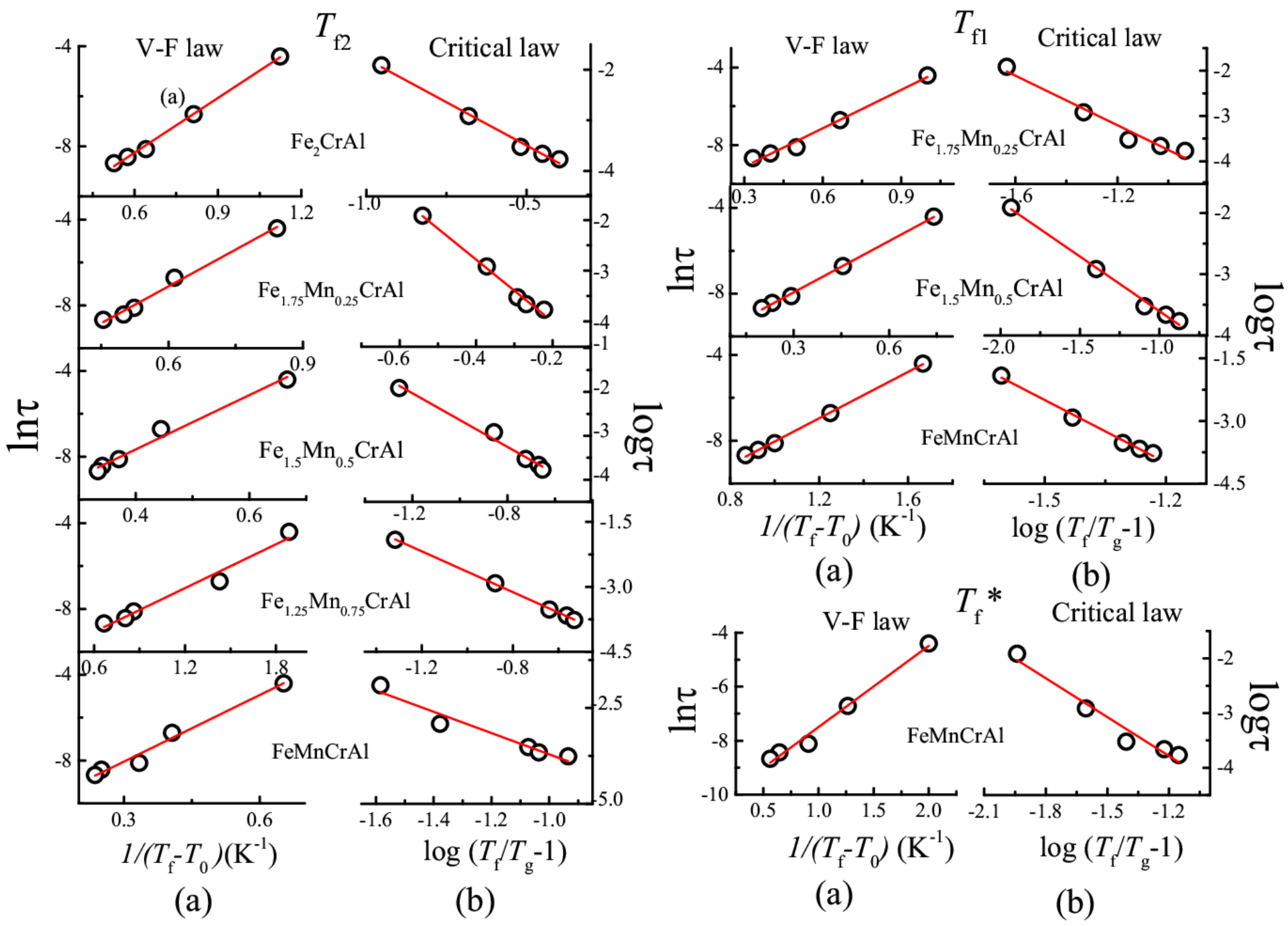

Figure 8 (a) V-F law fit of relaxation time (τ) as function of reduced temperature $1/(T_f\text{-}T_0)$ using eqn. 7 (b) Critical law fit of relaxation time (τ) as function of reduced temperature $(T_f/T_g\text{-}1)$ using eqn. 6 for $T_f^*$, $T_{f1}$ and $T_{f2}$ for respective alloys.

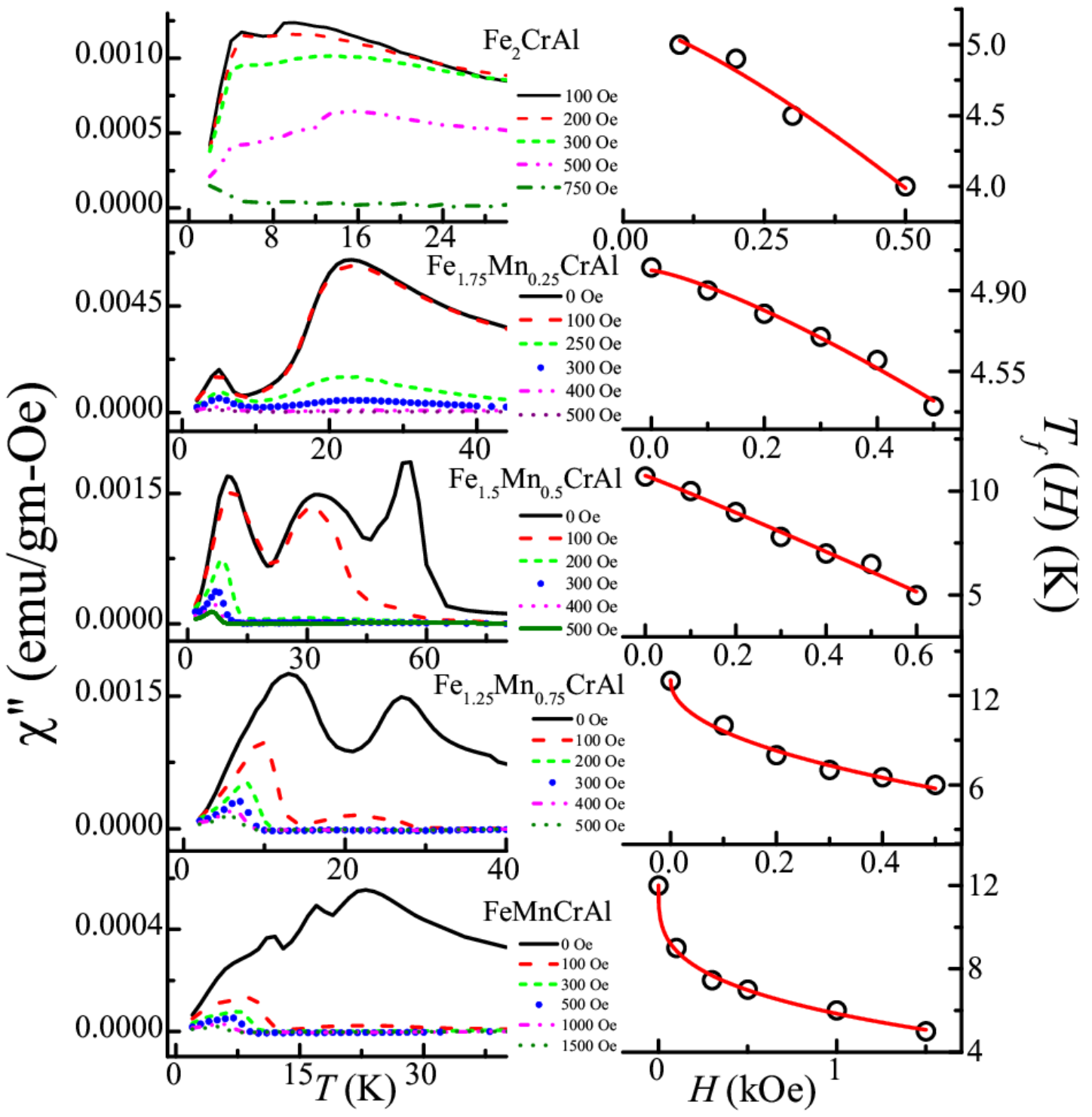

Figure 9 Left panel: Temperature response of out-of phase component of AC susceptibility at different $H_{DC}$. Right panel: $T_f$ vs. $H$ plot obtained from out-of-phase component of AC susceptibility and fitted (solid red line) with eqn. 8.

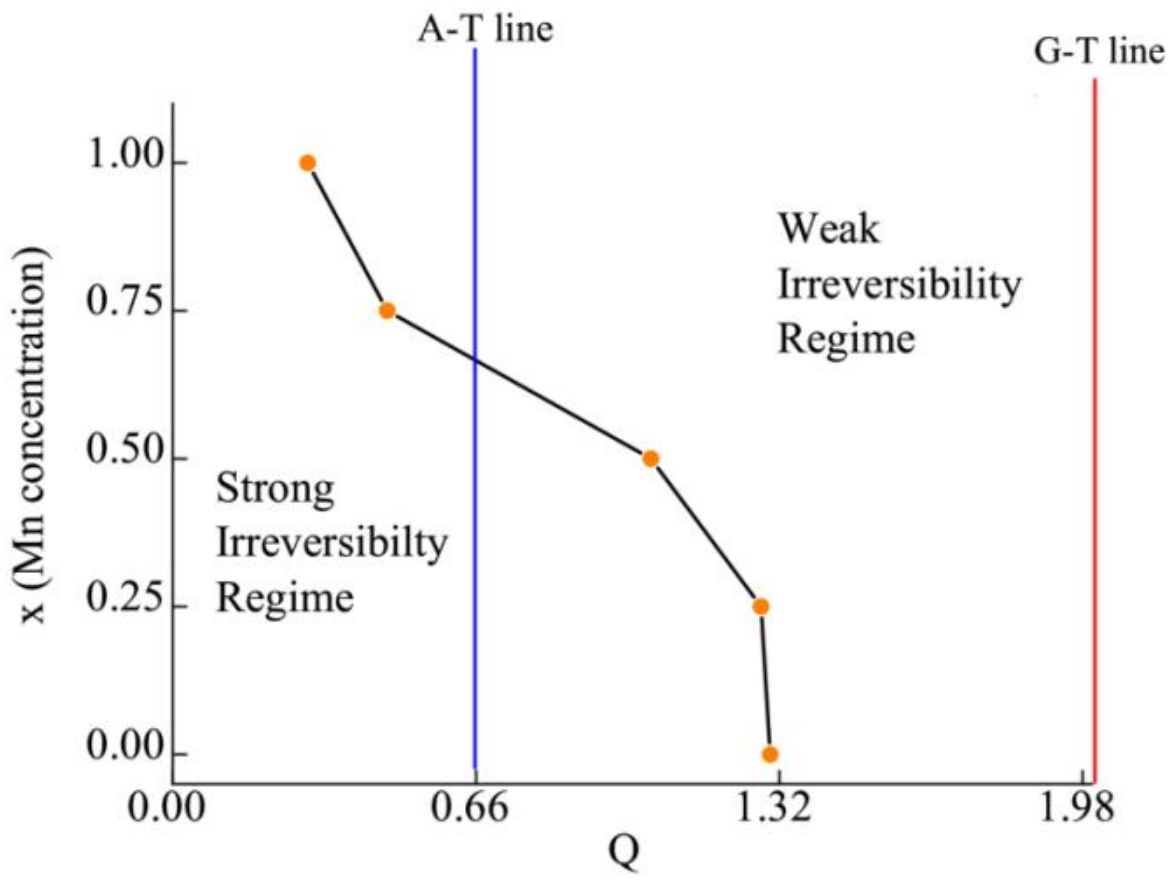

Figure 10 Phase diagram to illustrate the transition of the system from weak irreversibility regime to strong irreversibility regime.

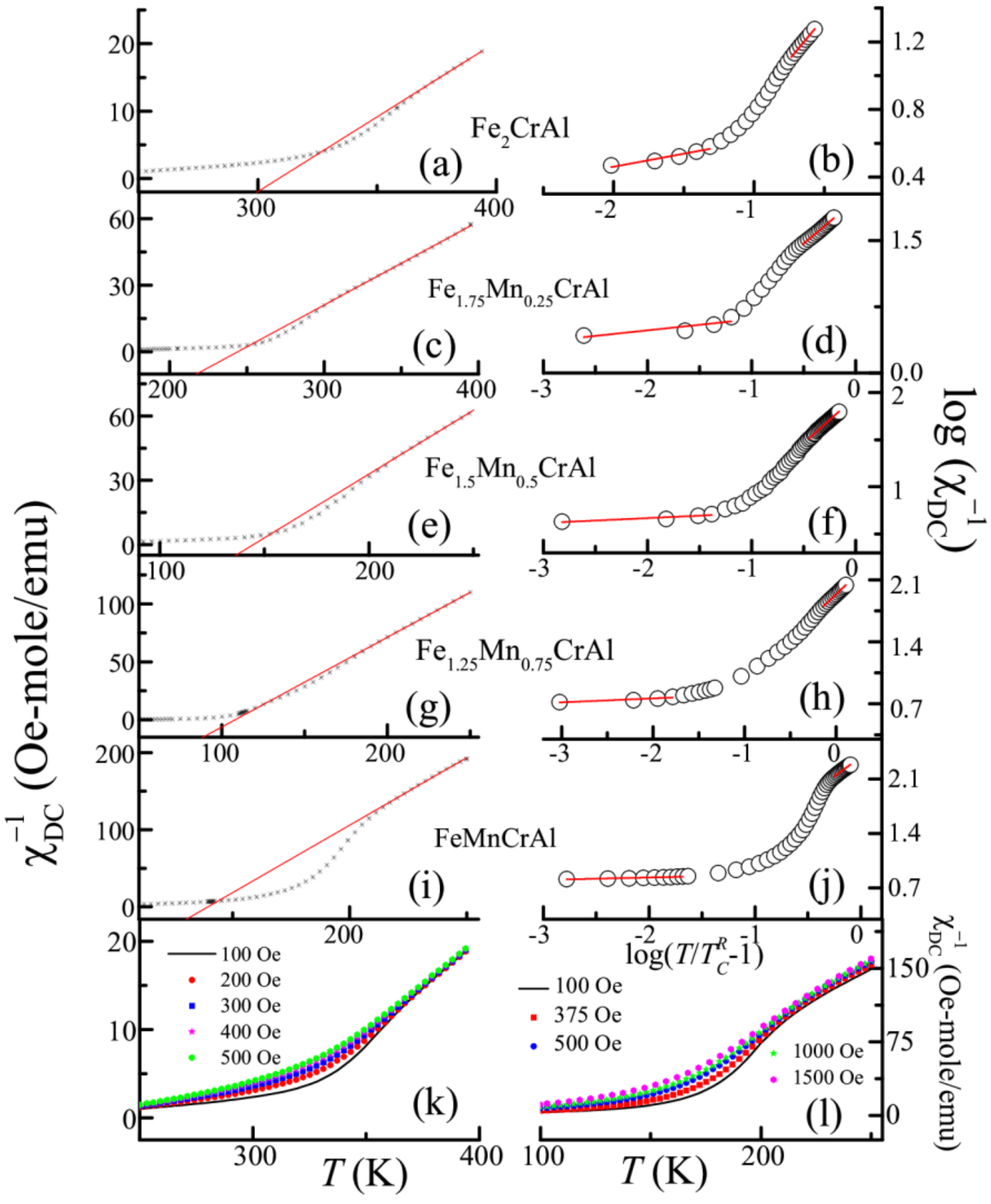

Figure 11 Left panel: (a), (c), (e), (g), and (i) represent Inverse DC susceptibility as function of temperature plot at 100 Oe. Solid red line represents the fitting using CW law. Right panel: (b), (d), (f), (h) and (j) represent the log-log plot of $\chi^{-1}_{DC}$ vs. $T/T_C^R$-1. Solid red lines are straight line fitting in GP and PM region. (k) and (l) depicts Inverse DC susceptibility as function of temperature plot at different applied fields for $Fe_2CrAl$ and FeMnCrAl respectively.

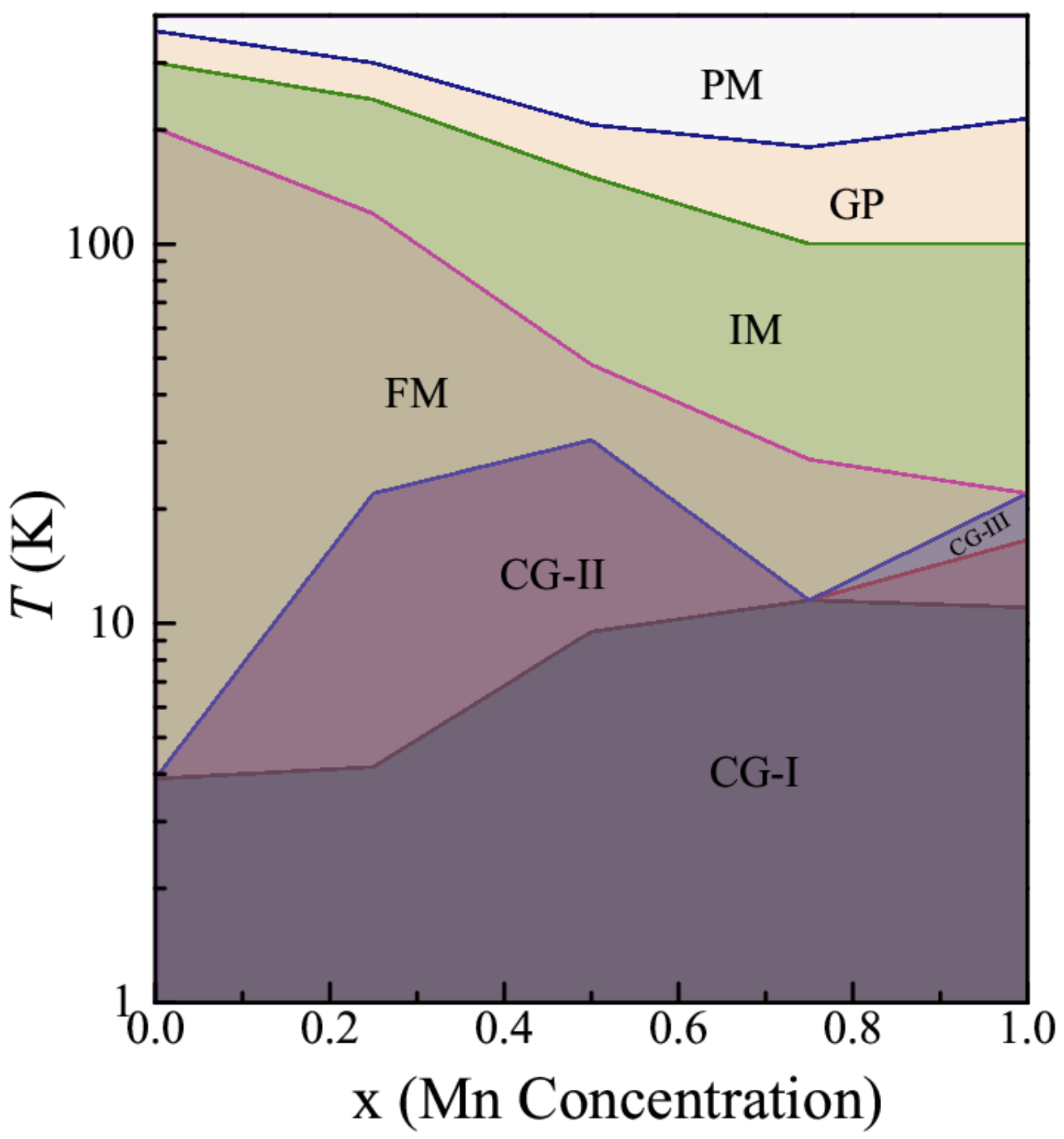

Figure 12 Temperature-concentration phase diagram of $Fe_{2-x}Mn_xCrAl$ ($0 \leq x \leq 1$).